\documentclass[]{raa}            % referee version: for submission
\usepackage{graphicx,times}
\usepackage{natbib}
\usepackage{epstopdf}
\usepackage{color}

\providecommand{\bm}[1]{\mbox{\boldmath$#1$\unboldmath}}

\def\xmm{{\sl XMM-Newton}}
\def\chandra{{\sl Chandra }}

\def\asca{{\sl ASCA }}
\def\sax{{\sl BeppoSAX}}
\def\rxte{{\sl RXTE}}

\def\ein{{\sl Einstein }}

\def\kmps{\hbox{km $\rm{s^{-1}}$}}

\def\fexxv{Fe~{\sc xxv}}
\def\fexxvi{Fe~{\sc xxvi}}

\def\fexx{Fe~{\sc xx}}

\def\nex{Ne~{\sc x}}

\def\ovi{O~{\sc vi}}

\def\mgxi{Mg~{\sc xi}}
\def\mgxii{Mg~{\sc xii}}

\def\six{Si~{\sc x}}

\def\sixiii{Si~{\sc xiii}}
\def\sixiv{Si~{\sc xiv}}

\def\sxv{S~{\sc xv}}
\def\sxvi{S~{\sc xvi}}
\def\arxviii{Ar~{\sc xviii}}

\begin{document}

   \title{Serendipitous Discovery of Warm Absorbers in the Seyfert 2 Galaxy IRAS 18325-5926
% $^*$
%\footnotetext{\small $*$ Supported by the National Natural Science Foundation of China.}
}

 \volnopage{ {\bf 2010} Vol.\ {\bf 9} No. {\bf XX}, 000--000}
   \setcounter{page}{1}

   \author{Shui-Nai Zhang
      \inst{1,2}
   \and Qiu-Sheng Gu
      \inst{1,2}
   \and Li Ji
      \inst{3}
   \and Zhi-Xin Peng
      \inst{1,2}
%   \and Chang-Sheng Shi
%     \inst{4}
   }

   \institute{Department of Astronomy, Nanjing University, Nanjing 210093, China; {\it snzhang@nju.edu.cn}\\
	  \and
		Key Laboratory of Modern Astronomy and Astrophysics, Nanjing University, Ministry of Education, Nanjing 210093, China\\
        \and
            Key Laboratory of Dark Matter and Space Astronomy, Purple Mountain Observatory, Chinese Academy of Sciences, Nanjing, 210008, China\\
%       \and
%            College of Material Science and Chemical Engineering, Hainan University, Hainan 570228, China\\ 
\vs \no
   {\small Received [2010] [09] [03]; accepted [2011] [05] [04] }
}

\abstract{ 
Warm absorption is a common phenomenon in Seyfert 1s and quasars, but rare in Seyfert 2s.
In this paper, we report the detection of warm absorbers with high energy resolution in the Seyfert 2 galaxy IRAS 18325-5926 for the first time with \chandra HETGS spectra.
An intrinsic absorbing line system with an outflow velocity $\sim400\,\rm{km\,s^{-1}}$ was found, which is contributed by two warm absorbers with FWHM of $570\rm{\,km\,s^{-1}}$ and $1360\rm{\,km\,s^{-1}}$, respectively.
The two absorbers were adjacent,  and doing a transverse motion across our line of sight.
We constrained the distance of the absorbers to a small value, suggesting that the absorbers may originate from the highly ionized accretion disk wind ejected 5 years ago.
The perspective of this type 2 Seyfert provides the best situation to investigate the vertical part of the funnel-like outflows.
Another weak absorbing line system with zero redshift was also detected, which could be due to Galactic absorption with very high temperature, or an intrinsic outflow with very high velocity $\sim6000\,\rm{km\,s^{-1}}$.
}

   \authorrunning{Zhang et al. }            %author_head in even pages
   \titlerunning{Warm Absorbing Gas in The Seyfert 2 Galaxy IRAS 18325-5926 }  % title_head in odd pages
   \maketitle

%% The author head (on even pages) and the title head (on odd pages) will be
%% automatically extracted from \author{} and \title{}. Whenever the title is too long,
%% you will be asked to supply a shorter one by inserting either \authorrunning{} or
%% \titlerunning{} before \maketitle. Anyway, you can specify your own heads.
%%
%%
%% Note: In the following text body of your manuscript, please note several differences from
%%       other major journals:
%% (1) \subsection{Please Capitalize the First Letter of Each Notional Word in Subsection Title}
%% (2) Please Capitalize the First Letter of Each Notional Word in all tables' captions

%
%________________________________________________ sections below
%
\section{Introduction}           %% first-level sections will be auto-capitalized
\label{sect:intro}
Warm absorber (WA) in the X-ray band was firstly found by \citet{halpern84} using \ein telescope, and has been studied extensively since then.
These ionized absorbing gas exists in about 50\% Seyfert 1s \citep{reynolds97, george98} and quasars \citep{piconcelli05, misawa07, ganguly08}, characterized by X-ray absorption lines which are usually blue-shifted by a few hundred \kmps.
As a consequence, WAs are considered as a potential form of Active Galactic Nuclei (AGN) feedback.
The fundamental question is how WAs originate and where they locate.
The proposed models include accretion disk wind \citep{elvis00}, clouds in the broad line region (BLR) \citep{risaliti10} or the narrow line region (NLR) \citep{kinkhabwala02}, winds from the putative obscuring torus \citep{krolik01}, and a shocked outflow \citep{pounds10}, which span a wide range in radial distance from the central ionizing source.
However, different models may predict similar spectra, e.g. NGC 3783 (\citealt{netzer03} vs \citealt{krongold05}) and NGC 4051(\citealt{krongold07} vs \citealt{steenbrugge09}).
Determination of the origin of WAs is still mired in ambiguity.

In Seyfert 2s, the direct view to the continuum source is blocked by the molecular torus, and the spectrum is featured by emission lines instead of warm absorptions \citep{sako02}.
Only a few possible absorption features have been reported in Seyfert 2s, such as NGC 3786 \citep{komossa97} and NGC 4507 \citep{comastri98,matt04}.
In this work, we searched through all type 2 AGNs in the archive of \chandra High Energy Transmission Grating Spectrometer (HETGS) observations, and focused specifically on investigating WAs in a serendipitous Seyfert 2 galaxy IRAS 18325-5926.
It has been observed with several early X-ray telescopes, such as \rxte, \sax, and \xmm, but none of which showed the ionized absorption \citep{iwasawa04}.
The luminosity distance of IRAS 18325-5926 is 81.1 Mpc based on the optical spectroscopic redshift 0.020 \citep{jones04}.
We adopt a cosmology of $\mathrm{H_0=73\,km\,s^{-1}\,Mpc^{-1}}$ throughout this paper.

 %or precisely Seyfert 1.9 \citep{iwasawa95}.

This paper is organized as follows. In section 2, we describe the data reduction. Section 3 is devoted to the analysis of the data, and section 4 to the detailed model of the ionized absorbers of IRAS 18325-5926. In section 5 we discuss our results and draw our conclusions in section 6. 

%% Authors can use \cite, \citep and \citet for citation.
%% You may also give a citation as 'Michel et al. 1992', and use Table~1 or Fig.~1
%% and so forth. Using \ref and \label for cross-references of Tables/Figures is
%% a good way in adjusting/adding/removing text, tables or figures.

\section{Observations and Data Reduction}
\label{sect:Obs}

\subsection{Data Reduction}
Two observations of IRAS 18325-5926 were performed with \chandra HETGS \citep{canizares05} in 2002 March (PI Canizares) with a total exposure time of 108 ks over 3 days (Table~\ref{tab:id}).
Data were reduced uniformly and in a standard way \citep{zhang11} using the \chandra Interactive Analysis of Observations (CIAO) software (Version 4.2) and the \chandra Calibration Database (Version 4.3.0).  
All measurements in the paper were obtained from the combined first-order spectra from both the High Energy Gratings (HEG) and the Medium Energy Gratings (MEG) instruments.
The statistic errors on the line energies are typically small, and likely dominated by the instrument uncertainties.
The relative wavelength accuracy of the HEG is 0.0010 \AA, and of the MEG is 0.0020 \AA.
The spectral analysis was carried out using the Interactive Spectral Interpretation System (ISIS version 1.6.3, \citealt{houck02}).
We adopted $C$-statistics \citep{cash76}  to find the best-fitting model parameters, and 90\% confidence for the quoted errors.

\begin{table}[htbp]
 \centering
  \begin{minipage}[]{70mm}
\caption[]{ \chandra HETGS observation log}
   \label{tab:id}
  \end{minipage}
  
   \begin{tabular}{cccc}
   \hline\noalign{\smallskip}
ID &  Start Time  & Exposure time \\
   \hline\noalign{\smallskip}
3148 & 2002.03.20 11:46:45 & 56.9 ks \\
3452 & 2002.03.23 16:11:16 & 51.1 ks \\
   \noalign{\smallskip}\hline
   \end{tabular}
\end{table}

\subsection{The Variations}
\begin{figure}[htbp] %  figure placement: here, top, bottom, or page
   \centering
       \includegraphics[angle=-90,width=70mm]{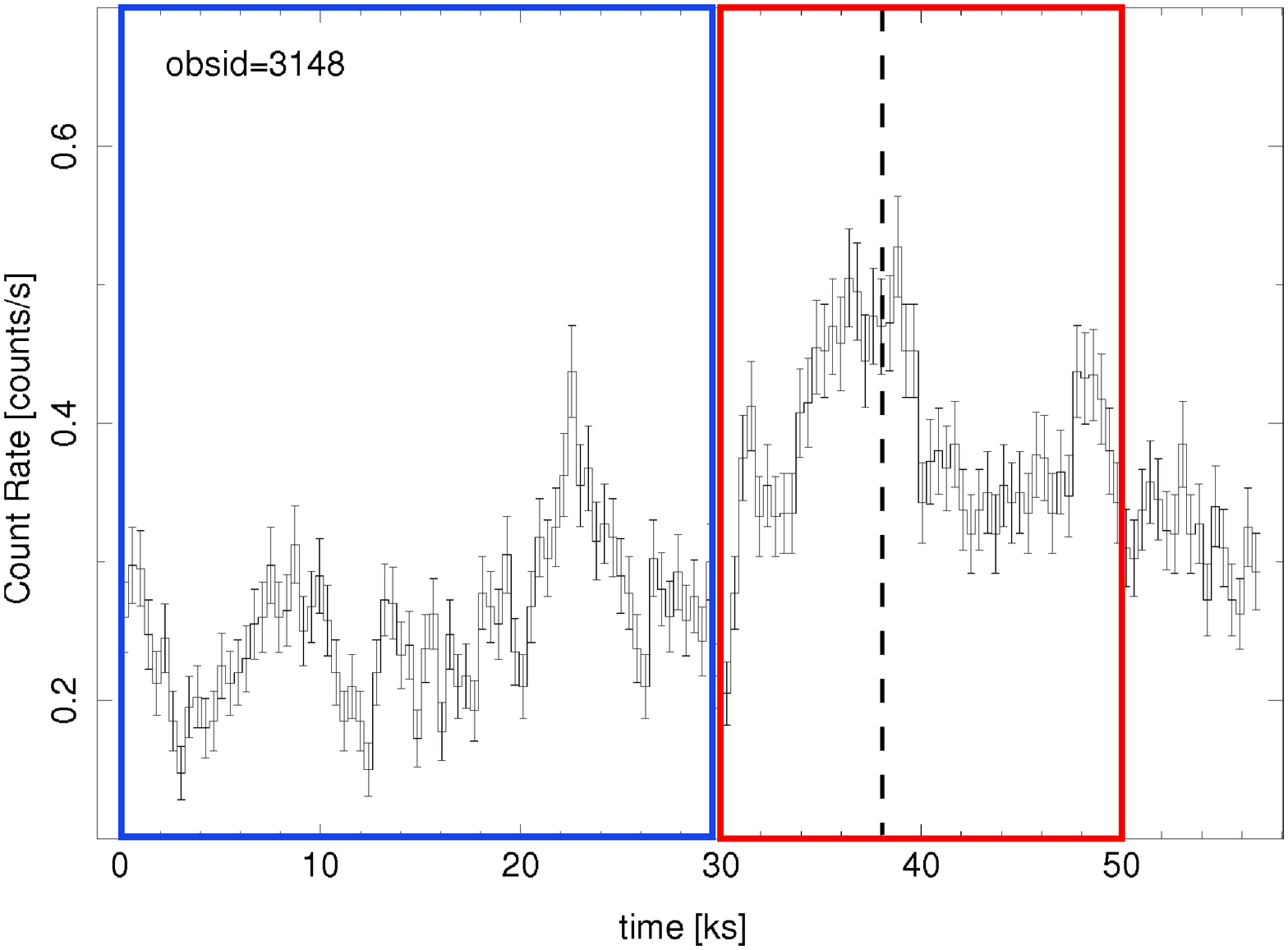} 
            \includegraphics[angle=-90,width=70mm]{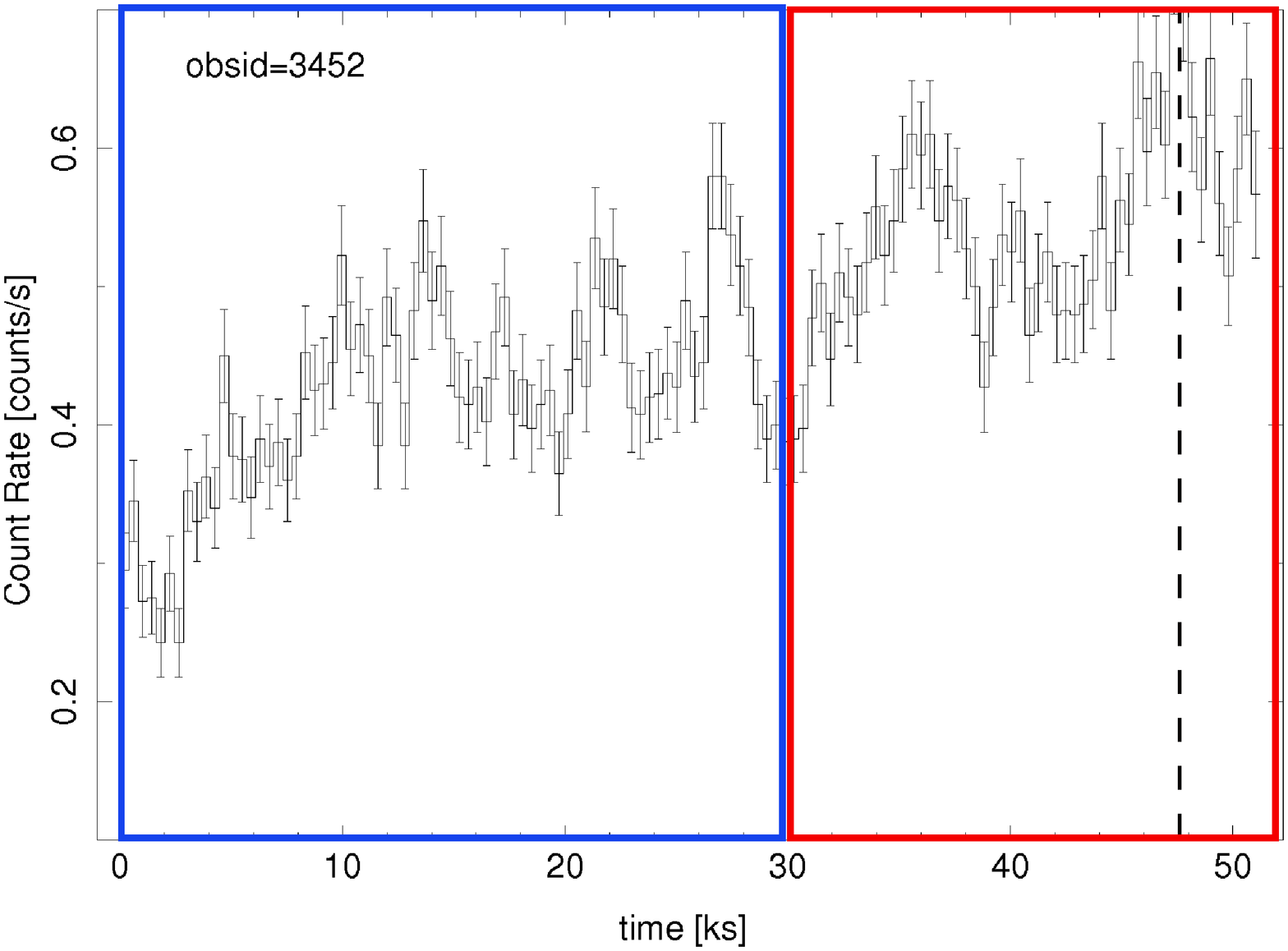} 
    \begin{minipage}[]{130mm}
   \caption{IRAS 18325-5926's X-ray light curves. The dash lines indicate two peaks, between which are 5 periodicities. The ``normal'' interval is marked with blue box, while the ``peak'' interval is marked with red box.}
   \label{fig:lc}\end{minipage}
\end{figure}

Fig.~\ref{fig:lc} shows the count rate of the short wavelength band (1.5-12 \AA) as a function of time for the two observations of IRAS 18325-5926, binned in intervals of 400 s.
The strong variations in a relatively short time scale are illustrated in the light curves.
The count rate varies from the minimum of 0.2 to the maximum of 0.6 during the observations.
We prepared two subsets of data according to the variation.

Our first group of data is observations of ID 3148 and ID 3452.
There is a 50\% change in the mean count rate between ID 3148 (0.3 counts per second) and ID 3452 (0.45 counts per second), that is separated by 3 days.
The large amplitude and the quick variability of the ionizing luminosity is helpful to determine the location of WAs, because one key diagnostic is based on applying non-equilibrium models on time variable AGN \citep{nicastro99}.

Our second group of data is ``normal'' and ``peak'', defined by the following strategy.
We marked the peaks of ID 3148 (0.5 counts per second) and ID 3452 (0.6 counts per second) with dash lines in Fig.~\ref{fig:lc},
and then marked off the ``peak'' interval (20 ks around the dash line, red part in Fig.~\ref{fig:lc}) and ``normal'' interval (30 ks before the ``peak'' interval, blue part in Fig.~\ref{fig:lc}) of each light curve.
The ``peak'' intervals of the two observations are combined and so do the ``normal'' intervals.
The mean count rate increased by a factor of 1.5 from ``normal'' to ``peak''.
A 58 ks periodicity of IRAS 18325-5926 was found in both a 5-day \asca observation \citep{iwasawa98} and totally 5-day \rxte~observations \citep{fabian98}.
The time interval between the two peaks of our light curves is 290 ks which happened to be 5 cycles, though the periodicity was not evident.
One of the potential mechanism for the periodicity is an object orbiting the black hole (BH) \citep{iwasawa98}.
Since this object could be the origin of WAs (e.g. the extension of a star), it's necessary to estimate the response in the opacity of WAs to the ionizing luminosity during a period.
``Normal'' and ``peak'' correspond to the phases where the object is moving beside and behind the BH.

\section{Spectral Analysis}

\subsection{Fitting the Continuum}
\begin{figure}[htbp] %  figure placement: here, top, bottom, or page
   \centering
      \includegraphics[angle=-90,width=\textwidth]{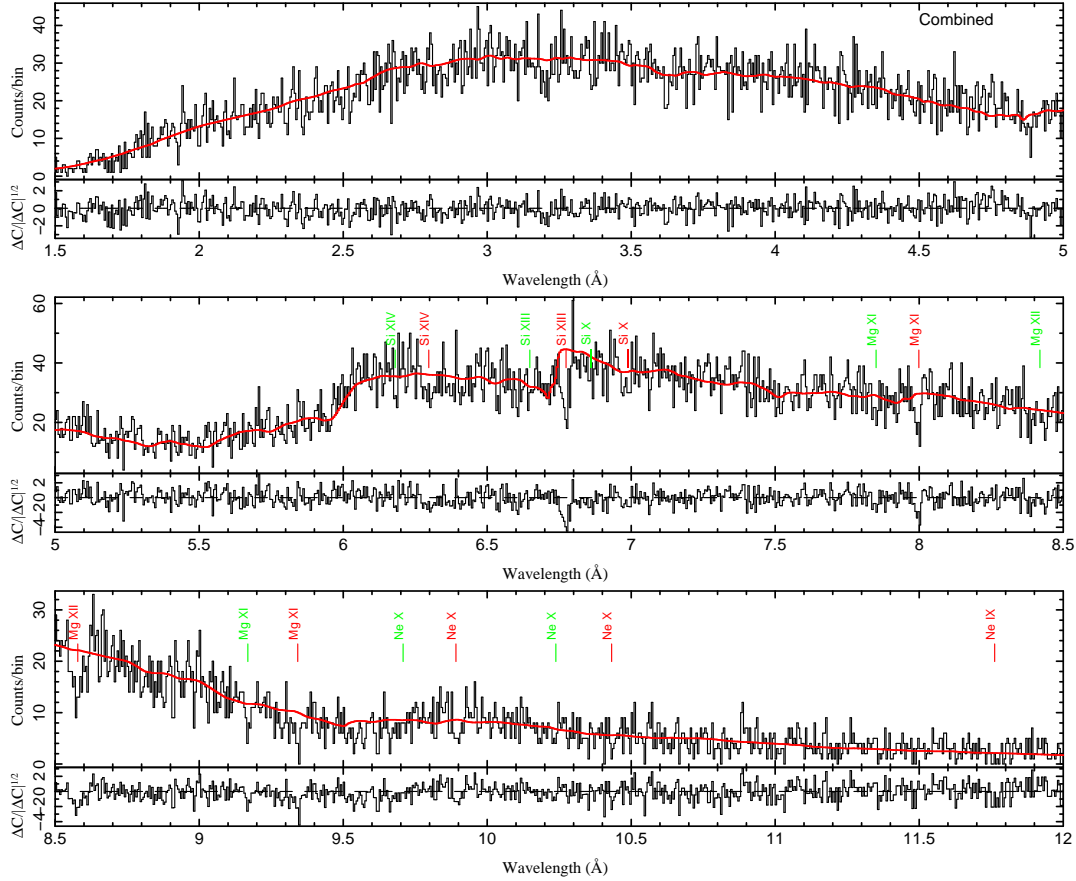} 
   \caption{IRAS 18325-5926's spectra. The intrinsic absorption lines are marked with red, while the absorption lines at zero redshift are marked with green.} 
       \label{fig:continua} 
\end{figure}

The IRAS 18325-5926 spectrum is affected by a relatively low Galactic absorption: $N_{H}=6.41\times10^{20}\,\mathrm{cm}^{-2}$ \citep{kalberla05}, which is included in the following analysis.
Since the soft X-ray band longer than 12 \AA~was heavily absorbed, we fitted the spectrum in 1.5-12 \AA~band using 0.005 \AA~wide bin.
All spectra were well fitted by an absorbed power-law except small residuals between 9.5 - 10 \AA~(Fig.~\ref{fig:continua}).
However, adding to the power-law continuum a thermal or a non-thermal component, does not improve the fitting statistics.
Table~\ref{Tab:conti} presents the values for the best continuum fit, indicating that the 2-10 keV flux of both groups of data increased by a factor of 1.5.
Neither the spectral slope nor the column density experienced significant variations.
%The comparison of the continuum models displays clear variations in each group of data, and the best fitting parameters are shown in Table~\ref{Tab:conti}.
The intrinsic neutral absorption is slightly under $10^{22}\,\mathrm{cm}^{-2}$, which is usually  the dividing line between type 1 and type 2 AGNs.
The variation of the absorbing column density is common for Serfert 2s, the time scale of which usually lasts for years \citep{risaliti02}, and several Seyfert 2s have been observed switching from Compton-thin to reflection-dominated or vice versa \citep{matt09}.
%The spectral slope of the combined spectrum is $\Gamma=1.89\pm0.02$, smaller than other X-ray telescope observations of IRAS 18325-5926 \citep{iwasawa04}.

%The black line is the combined spectrum, the red dashed line is the ID 3148 spectrum, the red dotted line is the ID 3452 spectrum, the blue dashed line is the ``normal'' spectrum, and the blue dotted line is the ``peak'' spectrum.
%The intrinsic column density is $N_{H}=0.96(\pm0.04)\times10^{22}\,\mathrm{cm}^{-2}$, and the spectral slope is $\Gamma=1.91\pm0.04$ ($\chi^2$/d.o.f.=1199/1047).
%The hard X-ray flux increased from 1.97 to $\rm{2.89\times10^{-11}\,erg\,s^{-1}\,cm^{-2}}$ by a factor of 1.5 between ID3148 and 3452. 
%The hard X-ray flux increased from 2.06 to $\rm{2.95\times10^{-11}\,erg\,s^{-1}\,cm^{-2}}$ by a factor of 1.5 between ``normal'' and ``peak''. 

%\begin{figure}[htbp] %  figure placement: here, top, bottom, or page
%   \centering
%   \includegraphics[angle=-90,width=\textwidth]{fig2.eps} 
%   \caption{The absorbed power-law model for continua of the combined spectrum and the two groups of data. The small valleys are absorption edges. The variation of the continua can be seen clearly.}
%   \label{fig:continua}
%\end{figure}

\begin{table}[htbp]
 \centering
 \begin{minipage}[]{90mm}
  \caption[]{Best-fitting parameters for the X-ray continua}
 \label{Tab:conti} \end{minipage}
  \begin{tabular}{cccccc}
  \hline\noalign{\smallskip}
State & $N_H$ & $\Gamma$& $\rm{Flux_{2-10keV}}$ & $\rm{L_{2-10keV}}$ &  $C$stat/d.o.f.  \\
 &\tiny  $(\rm{10^{22}\,cm^{-2}})$ &  & \tiny $\rm{(10^{-11}\,erg\,s^{-1}\,cm^{-2})}$  & \tiny $\rm{(10^{43}\,erg\,s^{-1})}$  &   \\
  \hline\noalign{\smallskip}
Combined & $0.90\pm0.04$ & $1.87\pm0.02$  & 2.41 &1.89 & 2378/2098  \\
ID 3148 & $0.94\pm0.06$ & $1.87\pm0.03$  &1.97 &1.55 & 2243/2098   \\
ID 3452 & $0.86\pm0.05$ & $1.88\pm0.03$ &2.89 & 2.28 & 2299/2098  \\
Normal & $0.89\pm0.06$ &  $1.90\pm0.03$ & 2.06 & 1.62 & 2232/2098   \\
Peak & $0.88\pm0.06$ & $1.83\pm0.03$ & 2.95 & 2.32 & 2287/2098  \\
 \noalign{\smallskip}\hline
  \end{tabular}
%%\vskip 20mm
%% place \tablecomments and \tablerefs below \end{center| and \end{center}:
%% you may leave the table-width parameter to editors or set to your actual size.
\end{table}

\subsection{Absorption Lines}
In order to increase the signal to noise ratio, the two observations were combined firstly for the detailed analysis.
We identified X-ray absorption in helium-like and hydrogen-like neon, magnesium, silicon, and L-shell silicon with -400 \kmps~of the line rest energy in IRAS 18325-5926 frame, marked in red in Fig.~\ref{fig:continua}. 
And then, we established another set of absorption lines including \mgxi(r) with Poisson probability of 0.9997 at redshift zero, or with -6000 \kmps~with respect to the systemic redshift, marked in green.

\begin{figure}[htbp] %  figure placement: here, top, bottom, or page
   \centering
   \includegraphics[angle=-90,width=\textwidth]{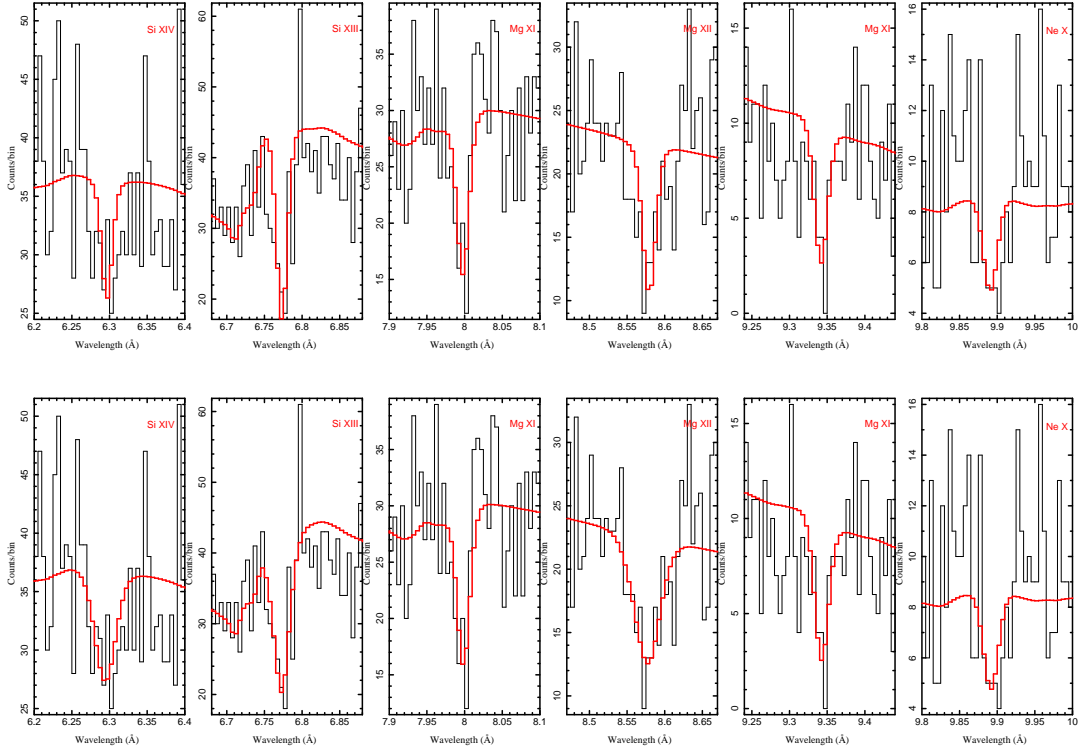} 
   \caption{Comparison of the best-fitting model of absorption lines. The lines in the upper panel is fitted with one set of Gaussian lines, while in the lower panel is fitted with two sets of Gaussian lines. Three highly ionized ions as \sixiii, \sixiv~and \mgxii~display broader Doppler width.}
   \label{fig:lines}
\end{figure}

The set of lines with -400 \kmps~velocity is similar to the absorption features found in Seyfert 1s, which are associated with X-ray WAs on the support of previous works \citep[e.g.][]{netzer03, mckernan07}.
Assuming that the lines came from a single WA, they should share the same outflow velocity and the same full width at half maximum (FWHM).
We used the two \mgxi~lines (\mgxi(r) \& \mgxi~Ly$\beta$) that are at the same atomic state to constrain the FWHM, and then applied it to the seven prominent lines (\sixiv, \sixiii, \six, \mgxi, \mgxii~and \nex) when fitting with Gaussian profiles.
Four lines of \six, \mgxi~and \nex~were well fitted with a FWHM of $\sim600\rm{\,km\,s^{-1}}$, but the other three highly ionized ions as \sixiii, \sixiv~and \mgxii~displayed broader Doppler widths, as shown in the upper panel of Fig.~\ref{fig:lines}.

\begin{figure}[htbp] %  figure placement: here, top, bottom, or page
   \centering
   \includegraphics[angle=-90,width=\textwidth]{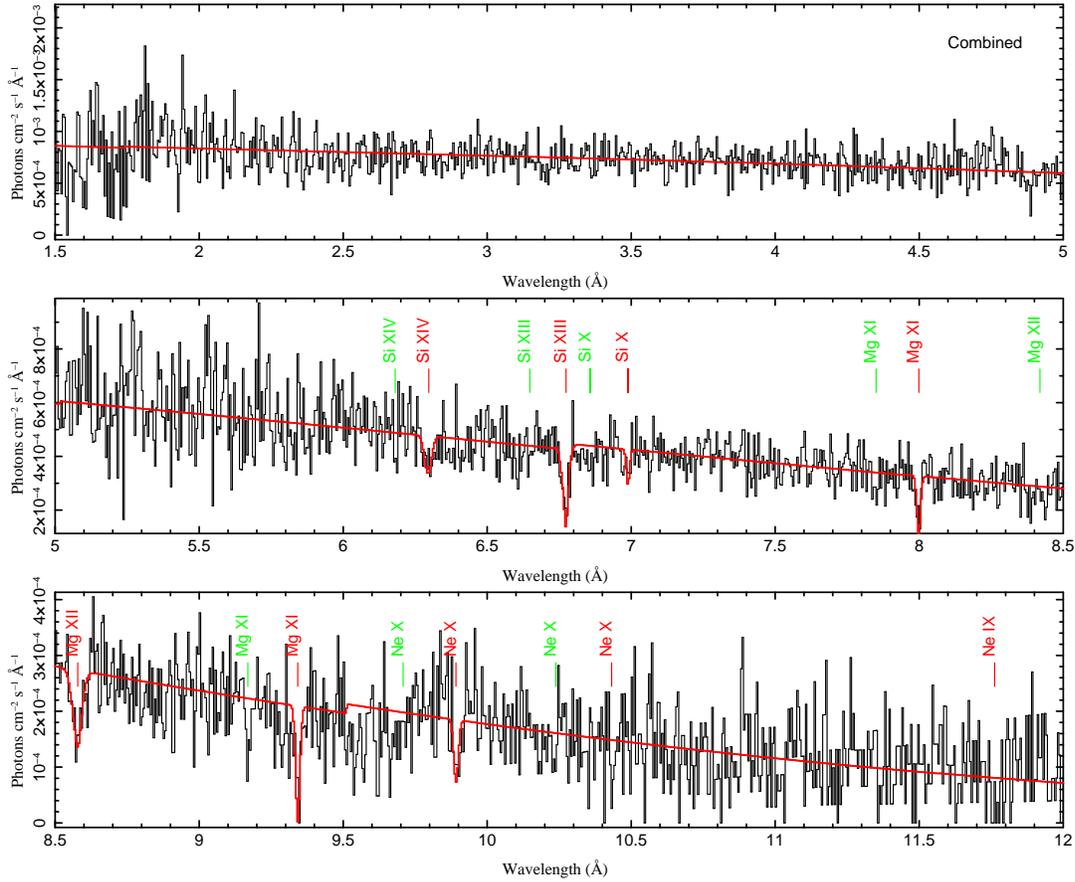} 
    %\begin{minipage}[]{80mm}  
   \caption{Comparison of the combined spectrum to the best-fitting model with two sets of Gaussian lines.}
   \label{fig:flux}
   %\end{minipage}
\end{figure}

As a consequence, we refitted the seven lines assuming they were from two WA components.
The refitting yielded an improvement to the three highly ionized lines ($\Delta C\sim6$ for 20 d.o.f. for each individual line) compared to the previous result, as shown in the lower panel of Fig.~\ref{fig:lines}.
These lines are decomposed into narrow and broad components, and the fitting parameters of dominant components are presented in Table~\ref{Tab:habs}.
The narrow component has the FWHM of $570\pm240\rm{\,km\,s^{-1}}$ and the outflow velocity of $340\pm110\rm{\,km\,s^{-1}}$,
while the broad component has the FWHM of $1360\pm560\rm{\,km\,s^{-1}}$ and outflow velocity of $460\pm220\rm{\,km\,s^{-1}}$.
The comparison of the combined spectrum to the best-fitting model is shown in Fig.~\ref{fig:flux}.

%though fitting with Gaussian profile usually drives the narrow component to small value.
%For ID 3148 data set, the FWHM=$580^{+400}_{-40}\rm{\,km\,s^{-1}}$ and the outflow velocity as $430\pm120\rm{\,km\,s^{-1}}$.
%One component is enough.
%For ID 3452 data set, there are two kind of models to fit the spectrum.
%One way is fitted with the broad component, while another way is with a narrow component.

%One absorption line with Poisson probability of 0.9997 at 9.169 \AA~was identified as \mgxi~at redshift zero.
Among the line system at zero shift, three more lines with Poisson probability greater than 0.990 other than \mgxi(r)~were found at 6.180 \AA, 7.850 \AA~and 8.438 \AA, which could be \sixiv, \mgxi~Ly$\beta$ and \mgxii, respectively.
One of the immediate explanation is the Galactic absorption.
Near the position of IRAS 18325-5926 ($l=336^{\circ}$, $b=-21^{\circ}$), there is a diffuse zero redshift ionized absorber traced by \ovi~in the UV band \citep[e.g.][]{savage03}.
The absorbing gas in the X-ray band may relate to this UV absorber, in the form of high velocity clouds in the Galactic halo \citep{fox06}.
High temperature would be necessary to produce such highly ionized ions. 
However, such a highly ionized Galactic absorption has never been reported before.

%Assistance is needed here, the temperature is about LogT=6.8 by line flux ratios. But not sure. Any reference book?}
Another possible interpretation is a high velocity outflow with 6000 \kmps.
The high velocity outflow is not rare in AGNs.
\citet{crenshaw99} performed a survey of Seyfert galaxies and found that nearly 60\% of the targets showed outflow velocities up to 2000 \kmps~in the UV band,
while in the X-ray band some 30\% of AGNs show evidence for an ionized wind of $v\sim0.1c$ though associated with \fexxv~and \fexxvi~lines most time \citep{tombesi10}.
We detected the absorption feature at 1.925 \AA, which could be \fexxv~with outflowing velocity of 13000 \kmps, however, its Poission probability is only 0.995.
Since the signal to noise ratio of this set of lines is not high enough, we can not tell between these two possibilities.

There are still some unidentified features.
Three absorption features around 3.2 \AA, 4.9 \AA~and 9.5 \AA~have Poisson probability as 0.9978, 0.9989 and 1.000, respectively.
The responsible ions could be \arxviii, \sxv~and \fexx~considering the systemic redshift, or these features could be spurious.
The emission-like feature at 9.85 \AA~is unidentified with Poisson probability of 0.0003.

\begin{table*}[htbp]
\centering
\begin{minipage}[]{100mm}
  \caption[]{Intrinsic absorption lines in the combined spectrum}
  \label{Tab:habs} \end{minipage}
  \begin{tabular}{lcccc}
  \hline\noalign{\smallskip}
Ion Name \& $\lambda_{rest}$ & FWHM  & $\lambda_{obs}$  & Flux & EWs  \\
 \tiny $\;\;\;\;\;\;\;\;$(\AA) &\tiny $\rm{km\,s^{-1}}$  & \tiny{(\AA)} & \tiny ($10^{-6}\,\mathrm{ph\,cm^{-2}\,s^{-1}})$ & \tiny (m\AA)   \\
  \hline\noalign{\smallskip}
\sixiv~Ly$\alpha$ (6.180)    &  1360 & 6.295 & $7.4\pm3.4$  &   $16\pm7$  \\
\sixiii(r) (6.648) &  ~ 570  &  6.773 & $\leq9.2$ & $\leq21$  \\
				  & 1360 & 6.771 & $9.0\pm6.2$ & $21\pm14$  \\
\six(r) (6.859)   & ~ 570 & 6.988 & $3.5\pm2.4$  & $8\pm6$  \\
\mgxi~Ly$\beta$ (7.850)  &  ~ 570 & 7.997 & $ 8.3\pm2.5$  & $26\pm8$  \\
\mgxii~Ly$\alpha$ (8.421) & 1360  & 8.579 & $12.7\pm5.2$ & $47\pm19$ \\
\mgxi(r) (9.169)  & ~ 570  &  9.341 & $ 13.8\pm3.8$  & $66\pm18$ \\
\nex~Ly$\gamma$ (9.708) & ~ 570  & 9.891 & $ 8.7\pm4.8$  & $48\pm26$  \\  
   \noalign{\smallskip}\hline   
  \end{tabular} 
\end{table*}

\section{Photoionization Model Fitting}
\label{sect:model}
In Sect. 3 we have identified the most prominent absorption lines.
In this Section we go further to fit the spectra with WA models.
There are good reasons for this. 
Firstly, the continuum absorption of the WA is not negligible, which is attributed to the neutral absorption in the previous section.
Then the contribution from the two WA components can be disentangled.
Finally, the predicted by model but very weak lines also help to constrain the spectrum.

\subsection{Photoionization Model}
The public available photoionization code XSTAR \footnote{http://heasarc.gsfc.nasa.gov/docs/software/xstar/xstar.html} is the tool to model the physical conditions of the absorbing gas. 
We use PVM\_XSTAR \citep{noble09} to generate grids of models, which provides parallel execution of XSTAR 2.1ln11.
The default solar abundances \citep{grevesse96} is adopted throughout this work.

\begin{figure}[htbp] %  figure placement: here, top, bottom, or page
   \centering
   \includegraphics[angle=0,width=100mm]{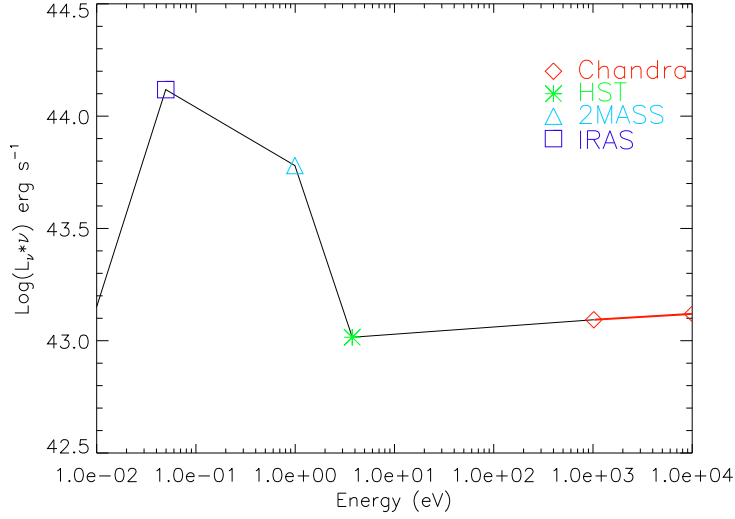} 
   %\begin{minipage}[]{60mm}
   \caption{The SED of IRAS 18325-5926. The red part is the intrinsic X-ray spectrum. The radio data point is beyond the left side of the figure.}
   \label{fig:sed}
   %\end{minipage}
\end{figure}

The observed spectral energy distribution (SED) is used as an input for generating the photoionization models.
We constructed the SED of IRAS 18325-5926 by including the intrinsic ionizing X-ray spectrum and obtaining some other data from NED:
4.85 GHz radio data from the Parkes-MIT-NRAO southern survey, 25 $\mu$m data from IRAS, J band data from 2MASS, and F330W data from HST.
The flux point at 1 keV band on the intrinsic X-ray continuum was simply linked to the optical point and to other data in the SED by straight lines in log-log space, as shown in Fig~\ref{fig:sed}.
The properties of WA model grids are mainly driven by the X-ray continuum of SED.
Removing a prominent IR-optical continuum bump from the SED yielded WA parameters that were within the 90 percent confidence intervals obtained when the bump was not removed \citep{mckernan03}, while the band below IR makes little difference to the ionization balance of WAs \citep{krolik01}.
Thus our SED is good enough to generate photoionization models for the two WA components with FWHM of 570 and $\rm{1360\,km\,s^{-1}}$.

%Assuming that the velocity dispersion is isotropic, i.e., $\sigma=\sqrt{3}\sigma_{line}$, and that $\sigma_{line}=\rm{FWHM}/2$, the velocity dispersion is $\frac{\sqrt{3}}{2}\rm{FWHM}$ (e.g., Netzer 1990, Netzer, H. 1990, in Active Galactic Nuclei, Saas-Fee Advanced Course 20, eds. T.J.-L. Courvoisier and M. Major, (Berlin: Springer–Verlag), p. 57).

The photoionization model used here contains three parameters: red-shift {\it z}; total neutral hydrogen column density $N_H$; and the ionization parameter
$\xi=L_{ion}/(n_eR^2)$, where $L_{ion}$ is the ionizing luminosity in the range 1-1000 Ryd, $n_e$  is the electron density and $R$ is the distance of the ionized gas to the central ionizing source.
$L_{ion}$ of IRAS 18325-5926 is about $\rm{1\times10^{44}\,erg\,s^{-1}}$, derived from the SED.

\subsection{The Warm Absorbers}
We fitted the combined spectrum and the two groups of data with the two photoionization models.
The results of our fit are summarized in Table~\ref{Tab:was}.

\begin{figure*}[htbp] %  figure placement: here, top, bottom, or page
   \centering
   \includegraphics[angle=-90,width=\textwidth]{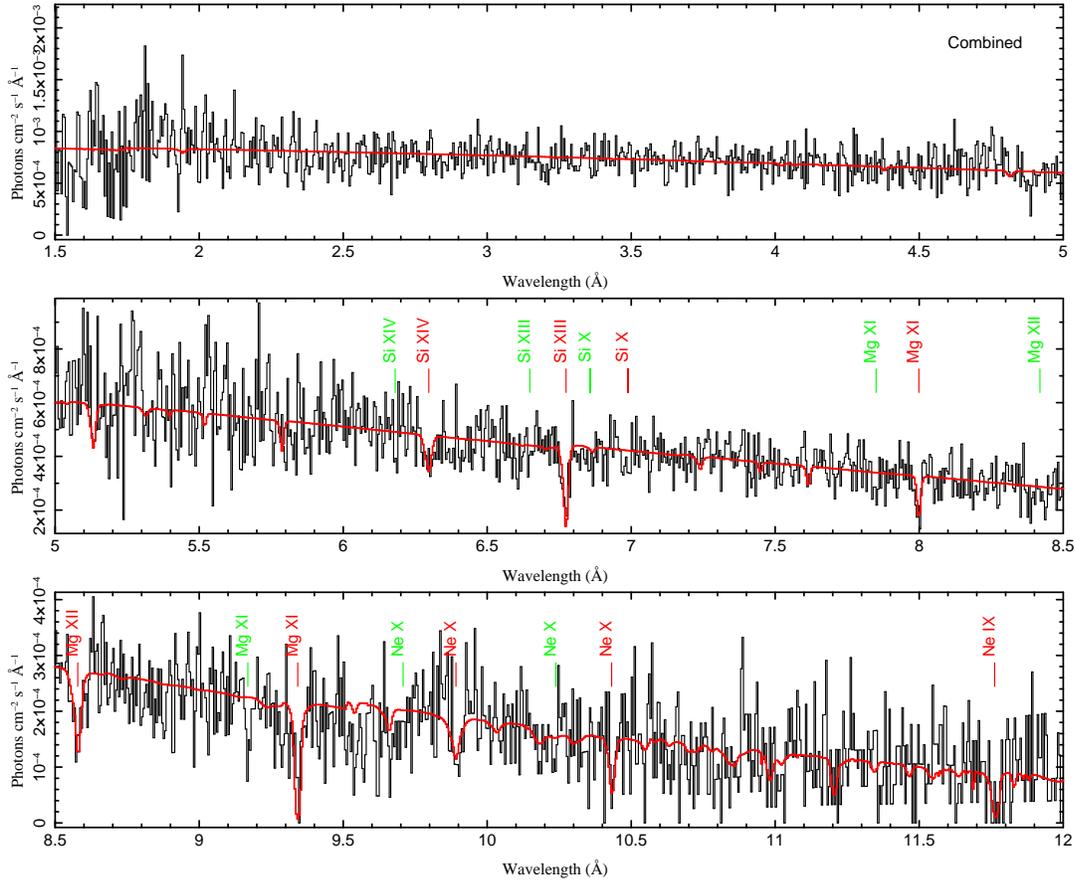} 
   \caption{ The combined spectrum fitted with the two WAs. Many weak absorption features are presented. The labels are the same as Fig.~\ref{fig:flux}.}
   \label{fig:waflux}
\end{figure*}

\begin{figure*}[htbp] %  figure placement: here, top, bottom, or page
   \centering
   \includegraphics[angle=-90,width=4.5in]{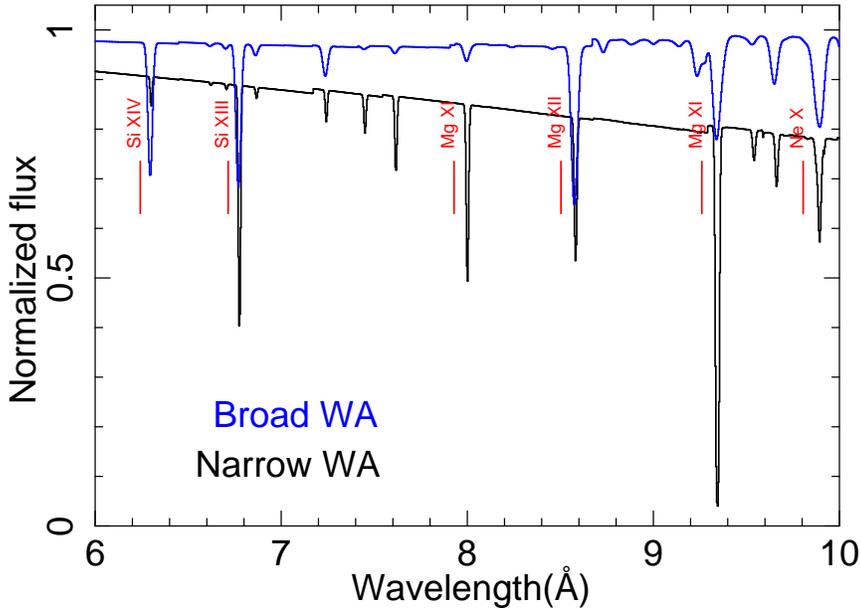} 
   \begin{minipage}[]{110mm}
   \caption{ The two WAs that are used to model the combined spectrum of IRAS 18325-5926 are shown on a scale where 1 is the incident continuum level. }
   \label{fig:model}
   \end{minipage}
\end{figure*}

For the combined spectrum, we show the best fitting model in Fig.~\ref{fig:waflux}, in which the absorption lines with -400 \kmps~are well represented.
The two photoionization WA models applied are displayed in Fig.~\ref{fig:model}, where the broad component contributes more in \sixiv, \sixiii~and \mgxii~lines but less in \mgxi~lines, and the narrow component also absorbs the continuum emission.
The broad WA is more ionized than the narrow WA, and seems to have a higher outflow velocity though the deviation is in the error tolerance.
Using the two photoionization models gives an improvement of $\Delta C=182$ compared to the continuum fitting.

%The total column density of the two WAs is about $\rm{1\times10^{22}\,cm^{-2}}$, which is equivalent to the contribution from cold column density of $\rm{0.2\times10^{22}\,cm^{-2}}$.

\begin{table}[htbp]
 \centering
 \begin{minipage}[]{90mm}
  \caption[]{Best-fitting parameters for the WA components}
  \label{Tab:was} \end{minipage}
  \begin{tabular}{ccc|ccc|cc}
  \hline\noalign{\smallskip}
State &cold $N_H$ & $\Gamma$& $N_H$ & Log$\xi$ & redshift $z$& $C$/d.o.f. & $^a\Delta C$   \\
 &\tiny  $(\rm{10^{22}\,cm^{-2}})$ &   &\tiny  $(\rm{10^{21}\,cm^{-2}})$& \tiny  (erg s $\rm{cm^{-1}}$) &   &  & \\
  \hline\noalign{\smallskip}
Combined & $0.74\pm0.09$ &   $1.94\pm0.02$   & $6.07\pm0.60$  & $1.58\pm0.09$ & $0.0190\pm0.0002$ & 2196/2092 & 182  \\  
          &                          &                              & $3.78\pm1.59$  & $2.35\pm0.25$ & $0.0184\pm0.0010$ &                      &   \\
  \hline\noalign{\smallskip}
ID 3148  & $0.71\pm0.07$ &   $1.95\pm0.03$   & $10.26\pm2.67$ & $\leq1.72$ & $0.0187\pm0.0003$ & 2160/2095 &  83 \\
  \hline\noalign{\smallskip}
ID 3452  & $0.71\pm0.10$ &   $1.95\pm0.04$   & $5.72^{+0.25}_{-2.09}$ & $1.58^{+0.02}_{-0.27}$ & $0.0192\pm0.0003$ &  2197/2092 & 102  \\
          &                          &                              & $4.56\pm1.92$ & $2.37\pm0.27$ & $0.0185\pm0.0010$ &                      &   \\
 \noalign{\smallskip}\hline
Normal  & $^b$0.71         &   $1.95\pm0.06$   & $7.55\pm0.98$ & $1.69\pm0.08$ & $0.0189\pm0.0002$ &  2207/2093 & 25  \\
%          &                          &                                & $^c$6.72          & $3.41\pm0.92$ & $^c$0.0188 &                      &   \\
 \noalign{\smallskip}\hline
Peak  & $^b$0.71           &   $1.94\pm0.03$   & $5.67^{+0.08}_{-1.12}$ & $1.41\pm0.17$ & $0.0194\pm0.0004$ &  2139/2093 & 148  \\
          &                          &                              & $4.86\pm2.07$ & $2.20\pm0.22$ & $0.0178\pm0.0009$ &                      &   \\
 \noalign{\smallskip}\hline
  \end{tabular}
\tablecomments{0.86\textwidth}{ $^a$ Change in the $Cash$ statistics between models with WAs and continuum models. $^b$ The column density is fixed at the value of the first group of data.}
  
\end{table}

\begin{figure*}[htbp] %  figure placement: here, top, bottom, or page
   \centering
   \includegraphics[angle=-90,width=\textwidth]{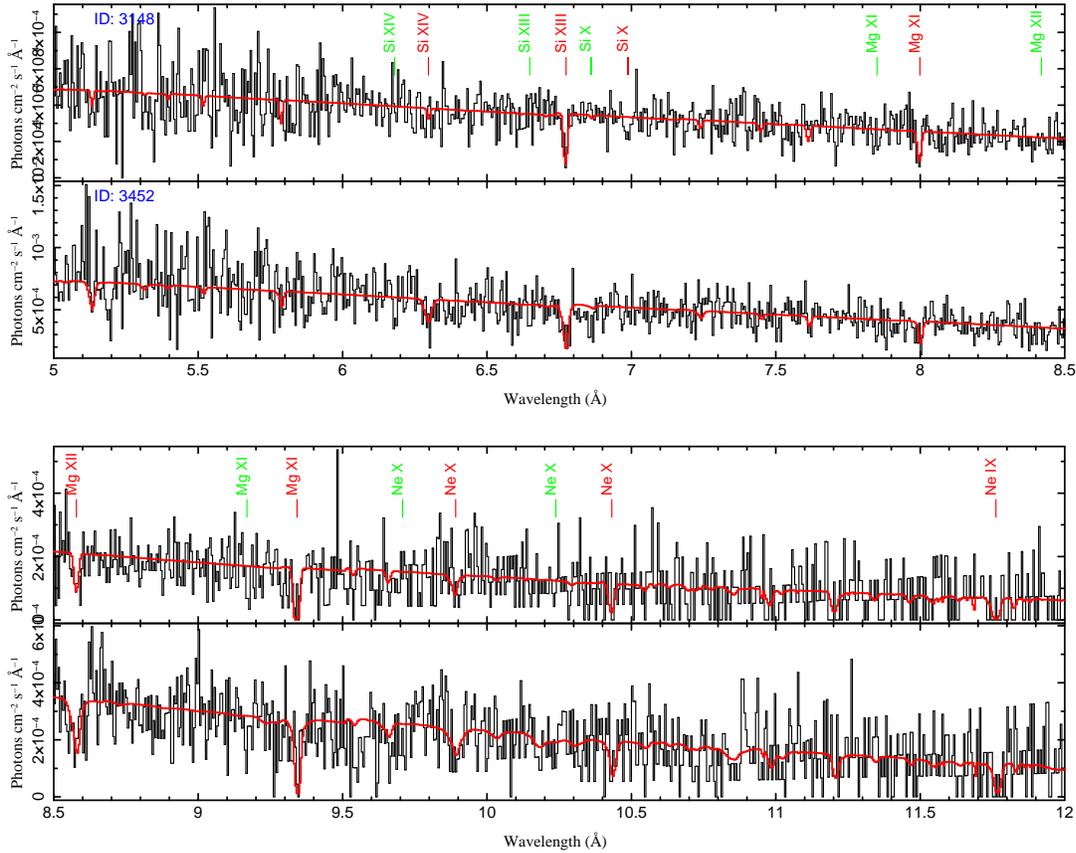} 
   \caption{ Comparison of ID 3148 and 3452 spectra fitted with photoionization models. \sixiii, \sixiv~and \mgxii~are broader in ID 3452 spectrum than that in ID 3148.}
   \label{fig:group1}
\end{figure*}

For the first group of data, we fitted ID 3148 spectrum with one narrow WA component, while for ID 3452 we used both the narrow and the broad WA components.
If we add a broad WA component to ID 3148, the column density of which would be only a few $\rm{10^{20}\,cm^{-2}}$ and the fitting does not give any improvements in the statistics.
The comparison of the spectra of ID 3148 and 3452 with their best fitting models is shown in Fig.~\ref{fig:group1}.
The ID 3148 spectrum has narrower \mgxii~and \sixiii~lines.
The cold gas column density are the same for ID 3148 to 3452, and the ionization parameter and redshift of narrow WA are within the error bars.
However, the column density of narrow WA decreased to a half from ID 3148 to 3452, and the broad WA showed up during 3 days.

For the second group of data, the cold gas column density was generally fixed to $N_H=0.71\,\rm{cm^{-2}}$ obtained from ID 3148 and 3452, which didn't vary in 3 days.
Like the first group, the ``peak'' spectrum contains a broad WA component, while the ``normal'' spectrum is well fitted with a single narrow WA.
Nothing changes significantly for the narrow WA, in particular the outflow velocity.
%Fitting with an additional highly ionized ($\xi\sim3.4$) broad component only gives an improvement of $\Delta C=3$ in the ``normal'' spectrum, and this counterintuitive result would stimulate vigorous discussion.

\begin{figure*}[htbp] %  figure placement: here, top, bottom, or page
   \centering
   \includegraphics[angle=-90,width=4.5in]{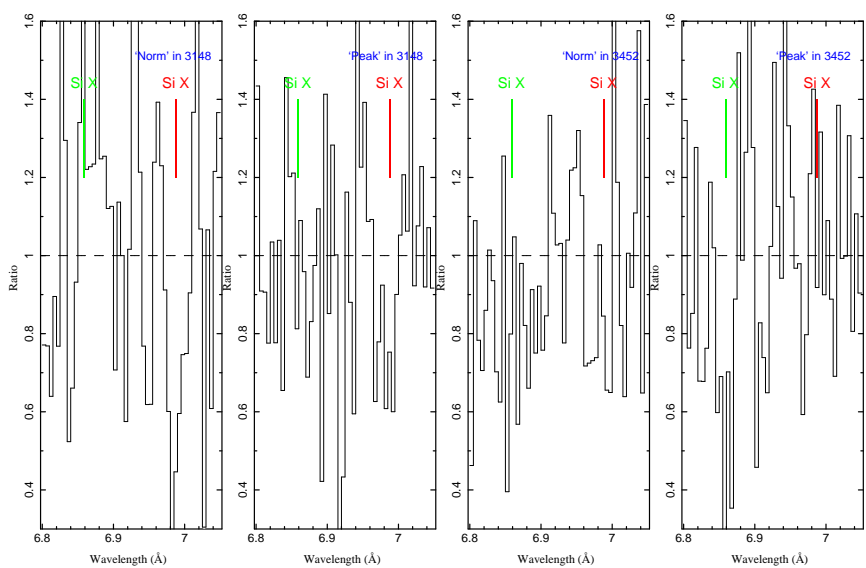} 
      \begin{minipage}[]{80mm}
   \caption{ The \six~lines in the different parts.}
   \label{fig:six}
     \end{minipage}
\end{figure*}

From above analysis the most interesting thing is that much stronger \six~absorption lines are displayed in both the ID 3148 and ``normal'' spectra.
Unfortunately, \six~absorption line would not be fitted with the photoionization models, because the silicon L-shell lines are omitted from the XSTAR database \citep{mckernan07}.
Thus we did a more detailed estimation for \six~line along the time sequence, as shown in Fig.~\ref{fig:six}.
It changed quickly responding to the luminosity: strongest to the lowest luminosity (``normal'' in ID 3148) but nearly disappeared to the highest luminosity (``peak'' in ID 3452).
The zero redshift \six~absorption line seems to change with luminosity too, getting stronger from the low luminosity to the high luminosity.
If this is confirmed, this set of lines would be determined as a high velocity outflow.

\begin{figure*}[htbp] %  figure placement: here, top, bottom, or page
   \centering
   \includegraphics[angle=-90,width=\textwidth]{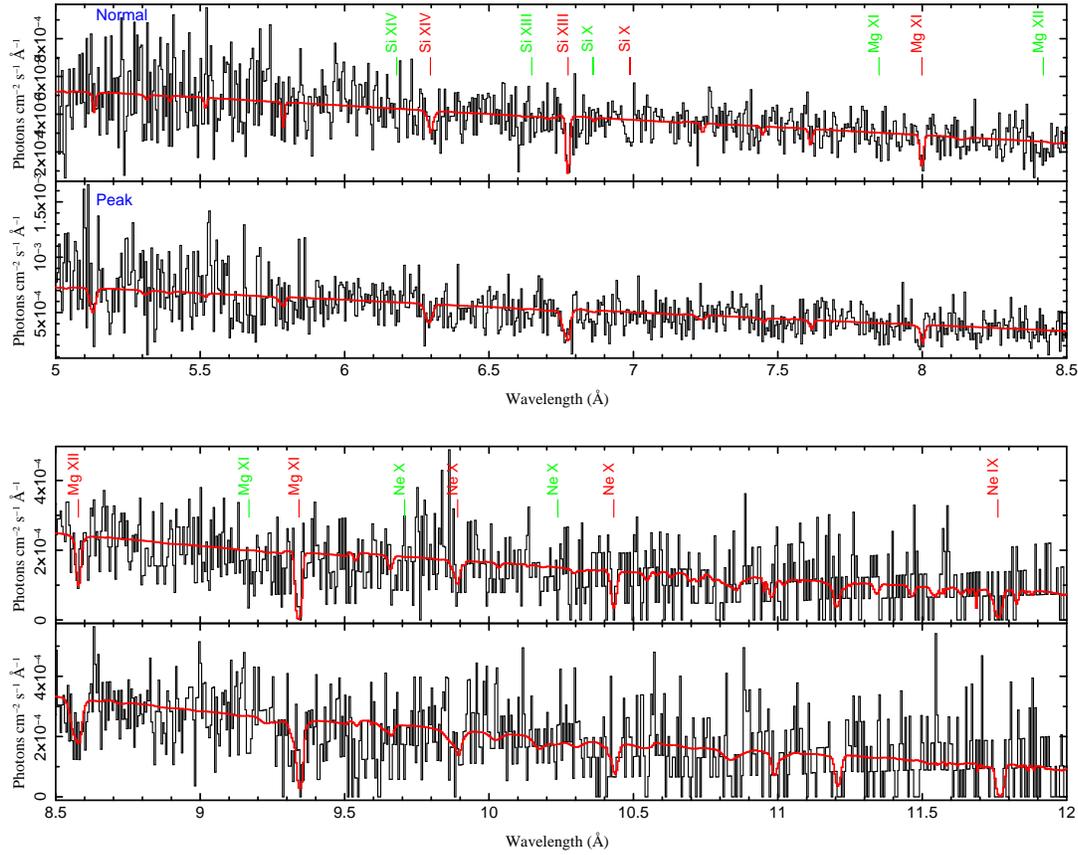} 
   \caption{ Comparison of ``normal'' and ``peak'' spectra fitted with photoionization models. \sixiii~and \mgxii~are broader in the ``peak'' spectrum, while \six~is much prominent in the  ``normal'' spectrum.}
   \label{fig:group2}
\end{figure*}

\section{Discussion}
\label{sect:discussion}

\subsection{The Location of WAs}
The periodicity of IRAS 18325-5926 didn't affect the WAs in a direct way.
There is no significant changes in the velocities between ``normal'' and ``peak'' states.
The existence of the broad WA in ``peak'' state is mainly due to the existence of it in ID 3452.
As a consequence, the option that a broad WA component is orbiting the BH is ruled out.

The large error bars in the ionization parameters of the narrow WA prevent us from detecting the response of WA to the ionizing luminosity.
However, the quick response in opacity of \six~which is lower ionized and thought to associated with the narrow WA indicates an electronic density larger than $5\times10^7\,\rm{cm^{-3}}$ \citep{nicastro99}.
Combined with the luminosity and the ionization parameter, the distance of the narrow WA from BH should be smaller than 0.07 pc.

We estimated the distance of the BLR to the BH as about 0.002 pc, using $r_{BLR}\propto L^{0.5}_{Bol,44}$ light days \citep{wandel99}.
On the other hand, the distance of the inner edge of the torus to the BH is about 1 pc, determined by $1\times L^{0.5}_{ion,44}$ pc \citep{krolik01}.
Considering $\xi=L_{ion}/(n_eR^2)$, for BLR and torus the electron density would be $2.5\times10^9\,\rm{cm^{-3}}$ and $5\times10^4\,\rm{cm^{-3}}$, respectively.
Since the column density of the broad WA is $5\times10^{21}\,\rm{cm^{-2}}$, the thickness of the clouds should be $1\times10^{-7}$ pc at BLR or 0.03 pc at torus.
A cloud near the torus with thickness of 0.03 pc, plugging in and covering our line of sight to the ionizing source in 3 days, seems counterintuitive.

As a consequence, combined with the constrain on distance by \six, the location of the WA is more likely near the BLR or accretion disk.
The broad WA that maybe a several million km thick cloud moved in to our line of sight, while half of the narrow WA moved out due to the decrease of the column density.

\subsection{The Origin of WAs}
Based on the \asca and \rxte~observations during 1997-1998 that is $4\sim5$ years before the \chandra observations, \citet{iwasawa04} found IRAS 18325-5926 had a steep continuum slope $\Gamma\sim2.2$ and prominent \fexxv~emission line (1.859 \AA) and \sxvi~radiative recombination continua (RRC, 3.548 \AA).
The authors suggested these features were caused by the reflection from a highly ionized disc.
%The X-ray spectrum above 2.0 keV is suggested to be including reflection from a highly ionized disc \citep{iwasawa04} based on a steep continuum slope $\Gamma\sim2.2$ and prominent \fexxv~emission line (1.859 \AA) and \sxvi~RRC (radiative recombination continua, 3.548 \AA) in the early observations $4\sim5$ years ago.
However, neither emission features can be detected in the \chandra observations, and the spectral slope is flatter than that in early observations. 
The reflection state is dissolved during the \chandra observations in 2002.

%\sxvi~RRC is even substituted by an absorption-like feature at 3.618 \AA~with Poisson probability of 0.9940.

It's interesting to consider where the highly ionized gas on the disc has gone.
If the gas travelled with a velocity of 400 \kmps~along our line of sight in the $4\sim5$ years, it would reach a projected distance of 0.002 pc.
The real distance of it to the BH is larger than or equal to 0.002 pc, which is quite coincident with the distance of WAs.
What's more, the adjacent gas regions in the ionized disk may have vastly different densities and temperatures \citep{nayakshin00}.
The large jump in temperature of the adjacent gas can explain the narrow and broad WAs naturally. 
The gas with FWHM of $570\rm{\,km\,s^{-1}}$ is followed by the gas of FWHM of $1360\rm{\,km\,s^{-1}}$, appeared as narrow and broad WAs.
When the narrow WA gradually passed by our line of sight, the followed broad WA just moved in.
As a result, the origin of the WAs should be the ionized wind from the accretion disc.

Since IRAS 18325-5926 is a Seyfert 2 galaxy, we are more likely seeing through the vertical part of the funnel-shaped thin shell outflow \citep{elvis00}.
In the vertical part, the transverse velocity is large, that is the reason of the quick change of the narrow WA to the broad WA.
Considering the highly ionized gas as the predecessor of the WAs, we predict that there is seldom WAs in IRAS 18325-5926 this year.

\section{Conclusions}
\label{sect:conclusion}
We draw a brief conclusion here.
We analyzed \chandra HETGS spectra of IRAS 18325-5926, and reported warm absorbers with high energy resolution in a Seyfert 2 galaxy for the first time.
An intrinsic absorbing line system with an outflow velocity $\sim400\,\rm{km\,s^{-1}}$ was found to be contributed by two warm absorbers, the FWHM of which are $570\rm{\,km\,s^{-1}}$ and $1360\rm{\,km\,s^{-1}}$, respectively.
The broad WA is only found in ID3452 observation and the column density of narrow WA decreased to a half from ID 3148 to 3452.
Thus we believe the two WAs were adjacent,  and doing a transverse motion across our line of sight.
We constrained the distance of the absorbers to a small value, and suggested that the absorbers came from the highly ionized accretion disk wind ejected 5 years ago.
The perspective of this type 2 Seyfert provides the best situation to investigate the vertical part of the funnel-like outflows.
Meanwhile, the periodicity of IRAS 18325-5926 didn't show any influence on the WAs at all.
Another set of possible zero redshift absorbing line system was also detected.
It could due to Galactic absorption with very high temperature, or an intrinsic outflow with very high velocity $\sim6000\,\rm{km\,s^{-1}}$.

\begin{acknowledgements}
The authors gratefully acknowledge the anonymous referee for their insights and comments that improved this paper.
The authors would also like to thank the support of the \chandra X-ray center.
Support for this work was provided by Program for New Century Excellent Talents in University (NCET), the National Natural Science Foundation of China under grants (10878010, 10221001 and10633040) and the National Basic Research Program (973 programme no. 2007CB815405). 
\end{acknowledgements}

\label{lastpage}

\end{document}